\documentclass[twocolumn,preprintnumbers,amsmath,amssymb]{revtex4}
\usepackage{graphicx}
\usepackage{dcolumn}
\usepackage{bm}

\begin{document}

\title{Morphogenesis by coupled regulatory networks}
\author{Thimo Rohlf}
\author{Stefan Bornholdt}
\affiliation{Interdisciplinary Center for Bioinformatics, University of Leipzig, 
Kreuzstra\ss e 7b, D-04102 Leipzig, Germany}
\date{\today}
\begin{abstract}
Based on a recently proposed non-equilibrium mechanism for spatial pattern formation 
\cite{RohlfBornholdt03} we study how morphogenesis can be controlled by locally 
coupled discrete dynamical networks, similar to gene regulation networks of cells in a 
developing multicellular organism. 
As an example we study the developmental problem of domain formation and proportion 
regulation in the presence of noise and cell flow. We find that networks that solve this task 
exhibit a hierarchical structure of information processing and are of similar complexity as 
developmental circuits of living cells. A further focus of this paper is a detailed study of 
noise-induced dynamics, which is a major ingredient of the control dynamics in the developmental network model. 
A master equation for domain boundary readjustments is formulated and solved for the 
continuum limit. Evidence for a first order phase transition in equilibrium domain size at 
vanishing noise is given by finite size scaling. A second order phase transition at increased 
cell flow is studied in a mean field approximation. Finally, we discuss potential applications. 
\end{abstract}
\maketitle

\section{\label{intro}Introduction}
Understanding the molecular machinery that regulates development of multicellular
organisms is among the most fascinating problems of modern science. Today, a growing 
experimental record about the regulatory mechanisms involved in development is 
accumulating, in particular in well-studied model-organisms as, e.g., \emph{Drosophila} 
or \emph{Hydra} \cite{TechnauHolstein,Bosch}. Still, the genomic details known today are not 
sufficient to derive dynamical models of developmental gene regulation processes in full detail.
Phenomenological models of developmental processes, on the other hand, are well established 
today. Pioneering work in this field was done by Turing, who in his seminal paper in 1952  
\cite{Turing52} considered a purely physico-chemical origin of biological pattern formation. 
His theory is based on an instability in a system of coupled reaction-diffusion equations. 
In this type of model, for certain parameter choices, stochastic fluctuations in the initial 
conditions can lead to self-organization and maintenance of spatial patterns, e.g.\ 
concentration gradients or periodic patterns. This principle has been successfully 
applied to biological morphogenesis in numerous applications 
\cite{GiererMeinhardt72,MeinhardtGierer2000}. However, as current experiments make 
us wonder about the astonishingly high complexity of single regulating genes in 
development \cite{BoschDevelGenes}, they also seem to suggest that diffusion models 
will not be able to capture all details of developmental regulation, and point at a complex 
network of regulating interactions instead. 

The role of information processing in gene regulatory networks during development has 
entered the focus of theoretical research only recently. One pioneering study was published 
by Jackson et al.\ \cite{Jackson86} in 1986, who investigated the dynamics of spatial pattern
formation in a system of locally coupled, identical dynamical networks. In this model, gene 
regulatory dynamics is approximated by Boolean networks with a subset of nodes 
communicating not only with nodes in the (intracellular) network, but also with some 
nodes in the neighboring cells. Boolean networks are minimal models of information 
processing in network structures and have been discussed as models of gene regulation 
since the end of the 1960s \cite{Kauffman1969,Origins}. The model of Jackson et al.\ 
demonstrated the enormous pattern forming potential of local information processing, 
similar to {\em direct contact induction} \cite{Slack93} as known in developmental biology.
More recently, Salazar-Ciuadad et al.\ introduced a gene network model based on 
continuous dynamics \cite{SalazarGarcia2000,SoleSalazar02} and coupling their 
networks by direct contact induction. Interestingly, they observe a larger variety of 
spatial patterns than Turing-type models with diffusive morphogens, and find that 
patterns are less sensitive to initial conditions, with more time-independent (stationary) 
patterns. This matches well the intuition that networks of regulators have the potential for 
more general dynamical mechanisms than diffusion driven models. 
Let us here therefore study interacting networks in pattern formation and in particular 
consider information-transfer-based processes. In \cite{RohlfBornholdt03}, we 
introduced a simple stochastic cellular automata model of spatial pattern formation
based on local information transfer, that performs de novo pattern formation by 
generating and regulating a domain boundary. In the following we will study how 
this very general mechanism could emerge as a result of interacting nodes in coupled 
identical networks, similar to gene regulation networks in interacting cells. 

The goal of this paper is twofold: First we will explore ways to map the earlier cellular 
automata model \cite{RohlfBornholdt03}, onto dynamical networks (Boolean networks 
and threshold networks), motivated by the general observation of gene regulation networks 
in biological development (section III). Secondly, we will give a more detailed account of 
the basic underlying mechanism, focusing on the stochastic dynamics (section IV) and 
on how the model was developed using a genetic algorithm (Appendix): In section IV, 
the statistical mechanics of noise-induced dynamics is studied in detail. Numerical 
evidence for a first order phase transition at vanishing noise rate is given. A master 
equation for boundary readjustments is formulated and solved in the continuum limit.
Robustness against cell flow is studied numerically and in a mean field approximation. 
In the Appendix, the genetic algorithm applied for identification of model solutions is 
explained and statistical properties of different solutions are compared. 

\section{Problem and Model}

Let us first motivate and define the morphogenetic problem that we will use as a test case. 
Then we undertake a three-step approach to find a genetic network model that solves this pattern formation 
problem. In the first step, we summarize the properties of the cellular automata model introduced in 
\cite{RohlfBornholdt03}. Cellular automata as dynamical systems discrete in time and state space 
are known to display a wide variety of complex patterns \cite{Wolfram84} and are 
capable of solving complex computational tasks, including universal computation.
We searched for solutions (i.e. rule tables which solve the problem) by aid of a genetic algorithm 
(for details, see Appendix). Candidate solutions have to fulfill four demands: Their update dynamics has 
to generate a spatial pattern which 1) obeys a predefined scaling ratio $\alpha/(1-\alpha)$, 2) is independent 
of the initial condition chosen at random, 3) is independent of the system size (i.e.\ the number of cells 
$N_C$) and 4) is stationary (a fixed point). In the second step, this cellular automata rule table is 
``translated'' into (spatially coupled) Boolean networks, using binary coding of the cellular automata states. 
The logical structure of the obtained network is reduced to a minimal form, 
and then, in step three, translated into a threshold network. 

\subsection{A test scenario: Hydra foot formation}
A classical model organism for studies of position dependent gene activation is the fresh 
water polyp \emph{Hydra}, which has three distinct body regions - a head with mouth and 
tentacles, a body column and a foot region. The positions of these regions are accurately 
regulated along the body axis. 

\begin{figure}[htb]
\let\picnaturalsize=N
\def\picsize{70mm}
\def\picfilename{fig1.eps}
\ifx\nopictures Y\else{\ifx\epsfloaded Y\else\input epsf \fi
\let\epsfloaded=Y
\centerline{\ifx\picnaturalsize N\epsfxsize \picsize\fi
\epsfbox{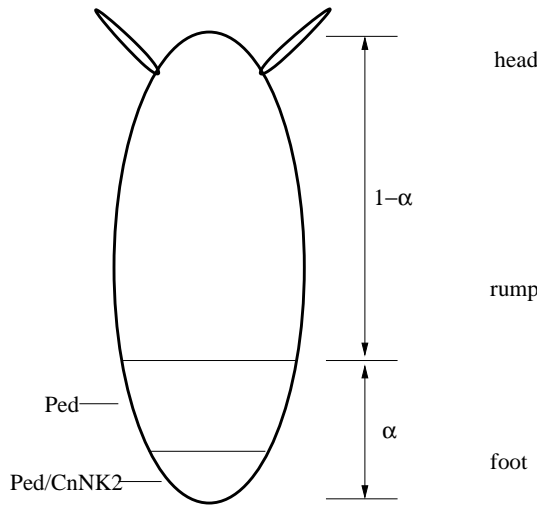}}}\fi
\caption{\small Gene expression domains in \emph{Hydra}, here for the example of the ``foot'' 
genes Pedibin (Ped) and Cn-NK2. The Pedibin domain shows a quite sharp boundary towards 
the body region of the animal. The relative position of the boundary, given by the ratio 
$\alpha/(1-\alpha)$, is independent of the absolute size of the animal (proportion regulation).}
\label{hydraped}
\end{figure}

The problem we shall focus on is sketched 
schematically in Fig. \ref{hydraped} for the example of the ``foot'' genes \emph{Pedibin} and 
\emph{CN-NK2} \cite{Thomsen2003}. CN-NK2 is active only in the lowest foot region,
the expression domain of Pedibin extends a bit more towards the rump region, where 
it is turned off at some point, i.e.\  the domain shows a sharp boundary. The relative 
position of the boundary is almost independent of the animal's size, i.e.\ the ratio 
$\alpha/(1-\alpha)$ (as denoted in Fig.\ \ref{hydraped}) is almost invariant under changes 
of body size. As the Pedibin/CN-NK2 system presumably plays an important role 
in determining the foot region, this invariance appears to be an essential prerequisite for 
maintaining the correct body proportions (proportion regulation) and to establish the 
head-foot polarity. Such regulation of position information is a quite general problem 
in biological development \cite{Wolpert1969}. 
An interesting problem is how the specific properties of this regulation 
can be achieved by a small network of regulatory genes and if so, whether local communication 
between the cells (networks) is sufficient. This basic question is the central motivation for the 
present study. In particular, we consider the simplified problem of regulating one domain 
as, for example, the foot region versus the rest of the body. We consider this as a one 
dimensional problem as first approximation to the well-defined head-foot-axis in \emph{Hydra}.  

\begin{figure}[htb]
\let\picnaturalsize=N
\def\picsize{85mm}
\def\picfilename{fig2.eps}
\ifx\nopictures Y\else{\ifx\epsfloaded Y\else\input epsf \fi
\let\epsfloaded=Y
\centerline{\ifx\picnaturalsize N\epsfxsize \picsize\fi
\epsfbox{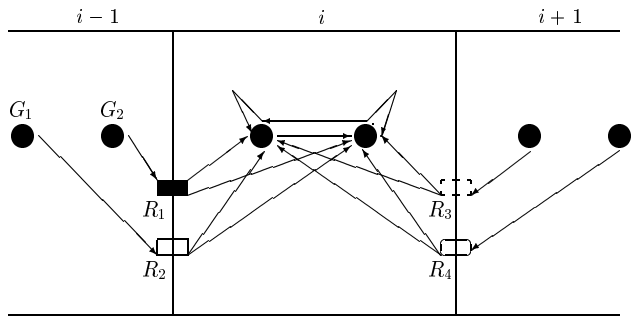}}}\fi
\caption{\small Diagram showing the interaction structure of the minimal network needed to solve 
the asymmetric expression task. For the sake of clarity, intracellular interactions between the two 
genes $G_1$ and $G_2$ are shown only for cell $i$, and likewise outgoing intercellular signals 
from the two genes two the neighbor cells $i-1$ and $i+1$ were left out. The transcription factors 
produced by gene $G_1$ and $G_2$ in cell $i-1$ couple to the receptor systems $R_1$ and $R_2$, 
respectively, whereas in cell $i+1$ the transcription factors produced by these genes couple to the 
receptor systems $R_3$ and $R_4$ (biased signaling). In cell $i$, the receptors release factors 
which regulate the activity of $G_1$ and $G_2$. }
\label{web1}
\end{figure}

Developmental processes exhibit an astonishing robustness. This often includes the ability of 
de novo pattern formation, e.g., to regenerate a
{\em Hydra} even after complete dissociation of the cell ensemble in a centrifuge \cite{Mueller}. 
Further, they are robust in the face of a steady cell flux: \emph{Hydra} cells constantly move from 
the central body region along the body axis towards 
the top and bottom, where they differentiate into the respective cell types according to their position 
on the head-foot axis. The global pattern of gene activity is maintained in this dynamic environment.  
Let us take these observations as a starting point for a detailed study how the interplay of noise-induced 
regulatory dynamics and cell flow may stabilize a developmental system.

\subsection{One dimensional cellular automata: Definitions}
To define a model system that performs the pattern formation task of domain self-organization 
\cite{RohlfBornholdt03}, consider a one-dimensional cellular automaton with parallel update
\cite{Wolfram84}. $N_C$ cells are arranged on a one-dimensional lattice, and each cell is labeled 
uniquely with an index $i \in \{0,1,...,N_C-1\}$.
Each cell can take $n$ possible states $\sigma_{i} \in \{0,1,..,n\}$. The state $\sigma_i(t)$ of cell $i$
is a function of its own state $\sigma_i(t-1)$ and of its neighbor's states  $\sigma_{i-1}(t-1)$  and
 $\sigma_{i+1}(t-1)$ at time $t-1$, i.e. 
\begin{equation}
\sigma_i(t) = f \left[ \sigma_{i-1}(t-1),  \sigma_i(t-1), \sigma_{i+1}(t-1) \right] 
\end{equation}
with $f: \{0,1,...,n\}^3  \mapsto \{0,1,...,n\}$ (a cellular automaton with {\em neighborhood 3}).
At the system boundaries, for simplicity we choose a discrete analogue of zero flux boundary 
conditions, i.e.\ we set $\sigma_{-1} = \sigma_{N_C+1} = const. = 0$. Other choices,
e.g.\ asymmetric boundaries with cell update depending only on the inner neighbor cell, lead
to similar results.
The state evolution of course strongly depends on the choice of $f$: for a three-state cellular automaton 
($n = 3$), there are $3^{27} \approx 7.626\cdot 10^{12}$ possible update rules, each of which
has a unique set of dynamical attractors. As we will show in the results section, $n=3$ is the minimal
number of states necessary to solve the pattern formation problem formulated above.

Now we can formulate the problem we intend to solve as follows: Find a set $\cal F$ of functions (update rules) 
which, given a random initial vector $\vec{\sigma} = (\sigma_0,..., \sigma_{N_c-1})$, within $T$ update steps 
evolves the system's dynamics to a fixed point attractor with the property:
\begin{equation}
\vec{\sigma}^* :=  \left\{  \begin{array}{c} \sigma_i = 2   \quad if \quad i < [\alpha\cdot N_C] \\ 
                                                \sigma_i \ne 2 \quad if \quad i \ge [\alpha\cdot N_C] \end{array}\right. 
\end{equation}
where $[ \cdot ] $ is the Gauss bracket. The scaling parameter $\alpha$ may take any value $0 < \alpha < 1$. For simplicity,
it is fixed here to $\alpha = 0.3$. Notice that $\alpha$ does not depend on $N_C$, i.e.\ we are looking for a set of solutions
where the ratio of the domain sizes $r:= \alpha / (1-\alpha)$ is {\em invariant under changes of the system size}.
This clearly is a non-trivial task when only local information transfer is allowed. 
The ratio $r$ is a {\em global property} of the system, which has to emerge from purely local (next neighbor) interactions
between the cell's states.

\subsection{Translation into spatially coupled Boolean networks}

One can now take a step further towards biological systems, by transferring the 
dynamics we found for a cellular automata chain onto cells in a line that communicate 
with each other, similar to biological cells. Identifying states with cell types and assuming 
that the model cells have a network of regulators inside, each of them capable to reproduce 
the rules of a cellular automaton, we obtain a model mimicking basic properties of a 
biological genetic network in development. 

Cellular automata rule tables can easily be translated into logical (Boolean) networks, e.g., for $n=3$, 
two internal nodes can be used for binary coding of the cell states. One then
has
\begin{equation}
(\sigma_{1}^i(t), \sigma_{2}^i(t))  = f_1(\sigma_{1,2}^{i-1}(t-1), \sigma_{1,2}^i(t-1), \sigma_{1,2}^{i+1}(t-1))
\end{equation}
with $f_1: \{0,1\}^{6} \mapsto \{0,1\}^2$.
The so obtained rule tables are, by application of Boolean logic,
transformed into a minimized \emph{conjunctive normal form}, which only makes use of the the
three logical operators NOT, AND and OR with a minimal number of AND operations. This is a rather 
realistic assumption for gene regulatory networks, as the AND operation is more difficult to realize on the basis
of interactions between transcription factors. Other logical
functions as, e.g., XOR, are even harder to realize biochemically \cite{Davidson01}. 
The structure of the constructed network and its biological interpretation is shown in Fig. \ref{web1}.
Note that the up-down symmetry of the body axis is broken locally by a spatially asymmetric receptor 
distribution \cite{Marciniac03}. 

\subsection{Translation into spatially coupled threshold networks}

Perhaps the simplest model for transcriptional regulation networks are threshold networks, 
a subset of Boolean networks, where logical functions are modeled by weighted sums of the 
nodes' input states plus a threshold $h$ \cite{Kuerten1988b,Kuerten1988a}. They
have proven to be valuable tools to address questions associated to the dynamics and 
evolution of gene regulatory networks 
\cite{Wagner1994,BornholdtSneppen2000,BornRohlf00,RohlfBorn02,RohlfBornMTBio}.

Any Boolean network which has been reduced to its minimized form (here, the conjunctive normal form),
 can be coded as a dynamical threshold network, with minimally
three hierachies of information processing (``input layer'': signals from
the neighbor cells at time $t-1$, ``hidden layer'': logical processing of the signals, ``output
layer'': states of the two ``pattern genes'' in cell $i$ at time $t$).  
The genes' states now may
take values $\sigma_i = \pm 1$, and likewise for the interaction weights one has $c_{ij}^l = \pm 1$ 
for activating and inhibiting regulation, respectively, and 
$c_{ij}^l = 0$ if gene i does not receive an input from gene $j$ in cell $l$.
The dynamics then is defined as
\begin{equation}
\sigma_{j}^i(t) = \mbox{sign}\,(f_j(t-1))
\end{equation}
with 
\begin{equation}
f_j(t)  = \sum_{k = 1}^2\sum_{l = i-1}^{i+1} c_{kj}^{l}\sigma_k^l + h_j 
\end{equation}
for the ``hidden'' genes, where $\sigma_k^l, k \in \{1,2\}$ is the state of the $k$th
\emph{output} gene in cell $l$ (there are no couplings between the genes in the ``hidden'' layer).
The threshold $h_j$ is given by
\begin{equation}
h_j = \sum_{k = 1}^2\sum_{l = i-1}^{i+1} |c_{kj}^l|-2,
\end{equation}
which implements a logical OR operation.
For the ``output'' (pattern forming) genes, one simply has 
\begin{equation}
f_k(t)  = \sum_{l = 1}^{k_{in}^k} \sigma_{l} - k_{in}^k,
\end{equation}
i.e.\ the weights are all set to one, and the (negative) threshold equals the number of inputs $k_{in}^k$ that 
gene $k$ receives from the ``hidden'' genes (logical AND).

\section{Results for deterministic dynamics}

\subsection{Cellular Automata Model}

\begin{figure}[htb]
\let\picnaturalsize=N
\def\picsize{70mm}
\def\picfilename{fig3.eps}
\ifx\nopictures Y\else{\ifx\epsfloaded Y\else\input epsf \fi
\let\epsfloaded=Y
\centerline{\ifx\picnaturalsize N\epsfxsize \picsize\fi
\epsfbox{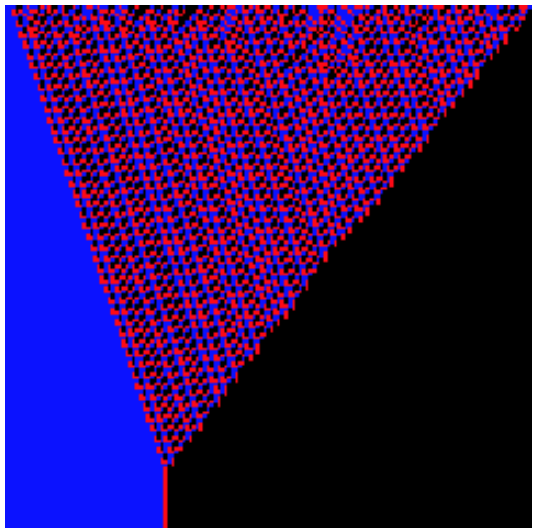}}}\fi
\caption{\small A typical dynamical run for the Automata as defined in Table I, here for a system of size $N_C = 250$
cells (deterministic dynamics, no noise), starting from a random initial configuration. Time is
running on the $y$-axis from top to bottom. Cells with state $\sigma_i = 0$ are depicted in black color, cells with
$\sigma_i = 1$ in red and cells with $\sigma_i = 2$ in blue.  }
\label{ca1g}
\end{figure}

\begin{figure}[htb]
\resizebox{85mm}{!}{\includegraphics{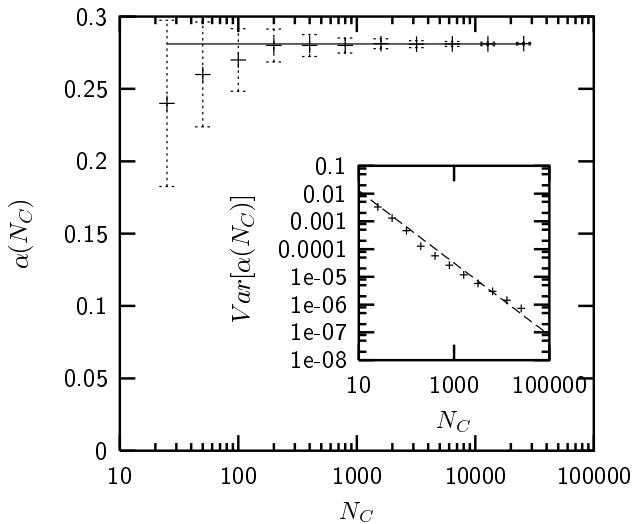}}
\caption{\small Finite size scaling of the self-organized relative domain size
$\alpha$ as a function of the total number of cells $N_C$. In the limit of large
system sizes, $\alpha$ converges towards a fixed value $\alpha_{\infty} = 0.281 \pm 0.001$
(as denoted by the straight line fit). The inset shows the finite size scaling
of the variance $Var[\alpha(N_C)]$; the straight line in this log-log plot has slope
$-1.3$ and indicates that fluctuations vanish with a power of the system size.}
\label{ndep}
\end{figure}

The first major outcome of the cellular automata model is that  a number of $n = 3$ different states is
necessary and sufficient for this class of systems to solve the given pattern
formation task. The update table of the fittest solution 
found during optimization runs, which solves the problem independent of 
system size for about 98 percent of (randomly chosen) initial conditions 
(i.e. has fitness $\Phi = 0.98$), is shown in Table I.
Fig. \ref{ca1g} shows the typical update dynamics of this solution. 
The finite size scaling of the self-organized relative domain size
$\alpha$ as a function of the the number of cells $N_C$ is shown in Fig. \ref{ndep}. In the limit of large
system sizes, $\alpha$ converges towards  
\begin{equation}
\alpha_{\infty} = 0.281 \pm 0.001.
\end{equation}
The variance of $\alpha$ vanishes with a power of $N_C$, i.e.\ the relative
size of fluctuations induced by different initial conditions becomes
arbitrarily small with increasing system size. Hence, the pattern
self-organization in this system exhibits considerable robustness against
fluctuations in the initial conditions. The main mechanism leading
to stabilization at $\alpha_{\infty} = 0.281$ is a modulation of
the traveling velocity of the right phase boundary in Fig.\ 2 
such that the boundary on average moves slightly
more than one cell to the left per update step, whereas the left boundary
moves one cell to the right exactly every third update step. 
the modulation of the right boundary can be seen as the result of interacting phase boundaries 
reminiscent of particle interactions. This picture of 
``particle computation'' is a useful concept also in various 
other contexts \cite{CompMechNew}.  
\begin{table}
\begin{tabular}{|c|ccc|c||c|ccc|c|}\hline
index & $\sigma_{i-1}$ & $\sigma_{i}$ & $\sigma_{i+1}$ & $\sigma_i$ & 
index & $\sigma_{i-1}$ & $\sigma_{i}$ & $\sigma_{i+1}$ & $\sigma_i$ \\ \hline
0 & 0 & 0 & 0 & 0 & 14 & 1 & 1 & 2 & 2\\ \hline
1 & 0 & 0 & 1 & 2 & 15 & 1 & 2 & 0 & 0 \\ \hline
2 & 0 & 0 & 2 & 1 & 16 & 1 & 2 & 1 & 0\\ \hline
3 & 0 & 1 & 0 & 0 & 17 & 1 & 2 & 2 & 1\\ \hline
4 & 0 & 1 & 1 & 2 & 18 & 2 & 0 & 0 & 0\\ \hline
5 & 0 & 1 & 2 & 2 & 19 & 2 & 0 & 1 & 0\\ \hline
6 & 0 & 2 & 0 & 1 & 20 & 2 & 0 & 2 & 0\\ \hline
7 & 0 & 2 & 1 & 2 & 21 & 2 & 1 & 0 & 1\\ \hline
8 & 0 & 2 & 2 & 2 & 22 & 2 & 1 & 1 & 2\\ \hline
9 & 1 & 0 & 0 & 0 & 23 & 2 & 1 & 2 & 2\\ \hline
10 & 1 & 0 & 1 & 1 & 24 & 2 & 2 & 0 & 1\\ \hline
11 & 1 & 0 & 2 & 1 & 25 & 2 & 2 & 1 & 2\\ \hline
12 & 1 & 1 & 0 & 0 & 26 & 2 & 2 & 2 & 2\\ \hline
13 & 1 & 1 & 1 & 1 &&&&&\\ \hline
\end{tabular}
\caption{\small Rule table of most successful best cellular automata solution found during genetic algorithm search. In the left column,
the rule table index is shown, running from 0 to 26, in the middle column the three input
states at time $t$ are shown, the right column shows the corresponding output states at time $t+1$.}
\end{table}
\begin{figure}[htb]
\let\picnaturalsize=N
\def\picsize{85mm}
\def\picfilename{fig5.eps}
\ifx\nopictures Y\else{\ifx\epsfloaded Y\else\input epsf \fi
\let\epsfloaded=Y
\centerline{\ifx\picnaturalsize N\epsfxsize \picsize\fi
\epsfbox{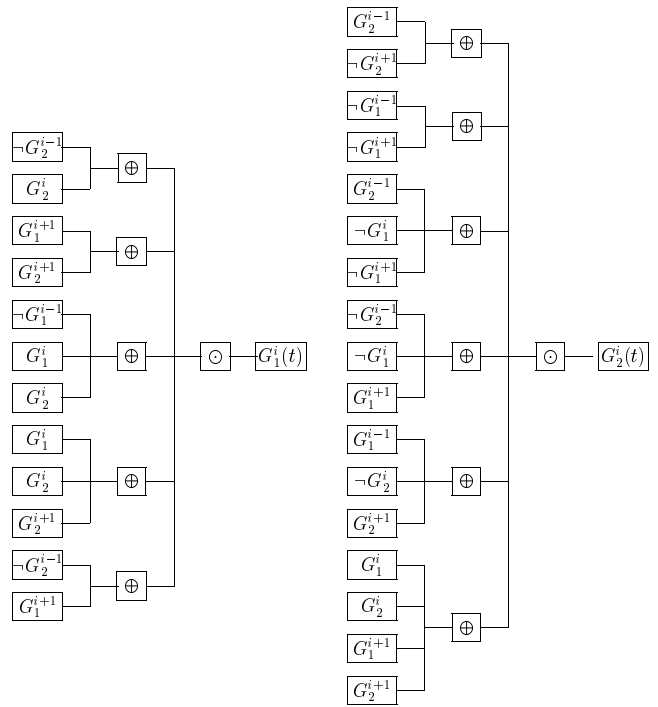}}}\fi
\caption{\small Boolean representation of the minimal network,
minimized conjunctive normal form. $G_a^b$
with $a \in \{1,2\}$ and $b \in \{i-1,i,i+1\}$
denotes gene $a$ in cell number $b$. The inputs in the
left branches of the trees are given by the genes'
states at time $t-1$. $\neg$ denotes NOT, $\odot$ denotes logical AND and $\oplus$ logical OR. }
\label{logBoolean}
\end{figure}

\subsection{Interaction topology of the minimal network}
In this section let us translate the rule table found in the previous 
section into a (gene) regulatory network. 
We will see that this network
has biologically realistic properties regarding the number
of genes necessary for information processing and the complexity
of interaction structure, making it well conceivable that
similar ``developmental modules'' exist in biological systems. 

\subsubsection{Boolean representation}

The rule table of Table 1 first is translated
into binary coding, i.e $0 \rightarrow 00$, $1 \rightarrow 01$
and $2 \rightarrow 10$, this corresponds to two ``genes'' 
$G_1$ and $G_2$ one of which ($G_1$) is active only in 
a domain at the left side of the cell chain. 
The so obtained Boolean update table is reduced to its minimized
conjunctive normal form, using a Quine-McCluskey
algorithm \cite{wutke}. For the construction of the network topology we use
the conjunctive normal form, as it is a somewhat biologically
plausible solution with a minimal number of logical AND operations.
In principle, other network topologies, e.g.\ with more levels of
hierarchy, are possible and biologically plausible, however, 
they involve a higher number of logical sub-processing steps, i.e.\ a
higher number of genes, hence we will not discuss them here.

Considering the huge number of possible input configurations which the outputs
theoretically could depend on, the complexity of the resulting network is surprisingly
low. As shown in Fig.\ \ref{logBoolean}, the output state of gene $G_1$ only depends
on five different input configurations of at maximum four different inputs,
gene number two on six different input configurations of at maximum four different inputs.
This indicates that the spatial information flowing into that network is strongly
reduced by internal information processing (only a small number of input states leads
to output ``1''), as expected for the simple stationary target pattern. Nevertheless,
this information processing is sufficient to solve the non-trivial task of domain
size scaling. 
\begin{figure}[htb]
\let\picnaturalsize=N
\def\picsize{85mm}
\def\picfilename{fig6.eps}
\ifx\nopictures Y\else{\ifx\epsfloaded Y\else\input epsf \fi
\let\epsfloaded=Y
\centerline{\ifx\picnaturalsize N\epsfxsize \picsize\fi
\epsfbox{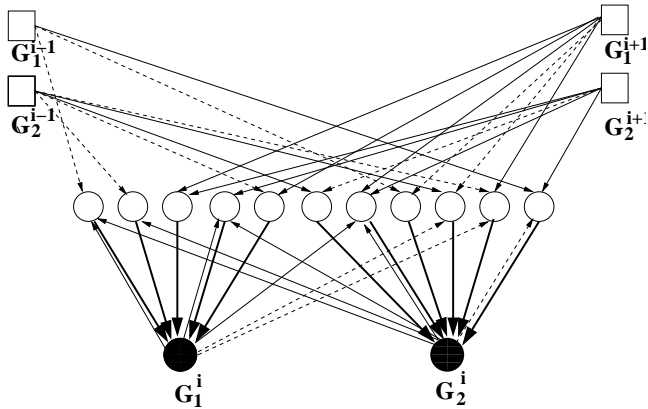}}}\fi
\caption{\small Threshold network realization of the pattern formation system. Solid
line arrows denote links with $w_{ij} = +1$, dashed arrows denote links with $w_{ij} = -1$.
The inputs from the genes in the neighbor cells ($G_1^{i-1}$,$G_2^{i-1}$,  $G_1^{i+1}$ and  $G_1^{i+1}$)
are processed by a layer of ``hidden genes'' (hollow circles) with different thresholds $h$ (as
these ``genes'' perform a logical OR operation, one has $h = k_{in}-2$). 
The threshold of gene $G_1$ is $h_1 = -5$, for gene $G_2$ one has $h_2 = -6$ (logical AND).}
\label{TNW}
\end{figure}

\subsubsection{Threshold network implementation}
Alternatively, the Boolean representation of the pattern formation system can be translated
into a three-layered threshold network (Fig.\ \ref{TNW}). The states of the genes $G_1$ and $G_2$
at time $t$ in a cell $i$ and its two neighbor cells $i-1$ and $i+1$
serve as inputs of 11 information-processing genes (``hidden'' layer). The state of these
genes then defines the state of $G_1$ and $G_2$ in cell $i$ at time $t+2$ (output layer).
Additionally, there is some feedback from $G_1$ and $G_2$ to the information processing layer,
as expected for the dependence on cell-internal dynamics already present in the cellular automata 
implementation of the model.
The resulting stationary spatial patterns of gene $G_1$ (the ``domain gene''), $G_2$
(active only at the domain boundary) and the ``hidden'' genes after system relaxation
are shown in Table 2, for a system size of $N_C = 30$ cells. 
From this seemingly redundant pattern in equilibrium one hardly guesses the higher 
level of genetic information processing during the self-organizing phase. 
The network we construct here, regarded as a ``developmental module'' defining the
head-foot polarity through spatially asymmetric gene expression, has a size similar to 
biological modules (compare, for example, the segment polarity network in \emph{Drosophila} \cite{vonDassow})
as well as similar complexity (average connectivity $\bar{K} \approx 3$).
\begin{table}
\begin{tabular}{|c|c|}\hline
gene number & spatial pattern  \\ \hline
1 & 111111111111111111111111111111 \\ \hline
2 & 111111111110111111111111111111 \\ \hline
3 & 111111111100000000000000000000 \\ \hline
4 & 111111111110000000000000000000 \\ \hline
5 & 111111111110111111111111111111 \\ \hline
6 & 111111111011111111111111111111 \\ \hline
7 & 111111111110000000000000000000 \\ \hline
8 & 100000000111111111111111111111 \\ \hline
9 & 000000000111111111111111111111 \\ \hline
10 & 111111111111111111111111111111 \\ \hline
11 & 111111111111111111111111111111 \\ \hline\hline
$G_1$ & 111111111100000000000000000000 \\ \hline
$G_2$ & 000000000010000000000000000000 \\ \hline
\end{tabular}
\caption{\small Spatial gene activity patterns of the network shown in Fig. 6. (stationary patterns after
the system has reached equilibrium), $N_C =30$. The numbers identify the ``hidden'' genes in Fig. \ref{TNW} from
the left hand side to the right. States were transformed by $\sigma \rightarrow (\sigma+1)/2$.}
\end{table}
\begin{figure}[htb]
\let\picnaturalsize=N
\def\picsize{80mm}
\def\picfilename{fig7.eps}
\ifx\nopictures Y\else{\ifx\epsfloaded Y\else\input epsf \fi
\let\epsfloaded=Y
\centerline{\ifx\picnaturalsize N\epsfxsize \picsize\fi
\epsfbox{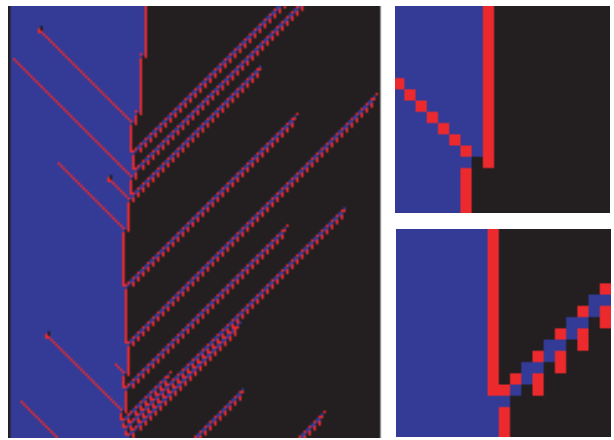}}}\fi
\caption{\small Quasi-particles, started by stochastic update errors, lead to control of the boundary
position under noise (left panel). The $\Gamma$ particle (top right) leads to readjustment of the boundary two cells to the left,
the $\Delta$ particle (bottom right) leads to readjustment of the boundary one cell to the right.}
\label{particles}
\end{figure}

\section{Dynamics under noise and cell flow}

In the following we will study dynamics and robustness of the model with respect to noise.
Two kinds of perturbations frequently occur: Stochastic update errors and external
forces cased by a directed \emph{cell flow} due to cell proliferations. Both types
of perturbations are very common during animal development, e.g., in  \emph{Hydra} cells continuously move from the central 
body region along the body axis towards the top and bottom, and differentiate into the respective
cell types along the way according to their position on the head-foot axis. 

Let us define stochastic update errors with probability $p$ per cell,
leading to an average error rate $r_e = p\,N_C$. Interestingly, 
this stochastic noise starts moving ``particle'' excitations in the cellular 
automaton which, as a result indeed stabilize the developmental structure of 
the system. To prepare for the details of these effects, define first how we 
measure the boundary position properly in the presence of noise.  
Let us use a statistical method to measure the boundary position
in order to get conclusive results also for high $p$: Starting at $i=0$,
we put a ``measuring frame'' of size $w$ over cell $i$ and the next $w-1$ cells,
move this frame to the right and, for each $i$,  measure
the fraction $z$ of cells with state $\sigma = 2$ within the frame. The algorithm
stops when $z$ drops below $1/2$ and the boundary position is defined to be $i + w/2$.

It is easy to see that, for not too high $p$, there are only
two different quasi-particles (i.e. state perturbations
moving through the homogeneous phases), as shown in Fig. \ref{particles}. In the following,
these particles are called $\Gamma$ and $\Delta$. The  $\Gamma$
particle is started in the $\sigma_2$ phase by a stochastic error $\sigma_i = 2
 \rightarrow \sigma_i \ne 2$ at some $i < \alpha N_C$), moves to the right and,
when reaching the domain boundary, readjusts it
two cells to the left of its original position. The $\Delta$ particle is started in the $\sigma_0$
phase by a stochastic error $\sigma_i = 0 \rightarrow \sigma_i \ne 0$ at
some $i > \alpha N_C$ and moves to the left. Interaction with the domain
boundary readjusts it one cell to the right. Thus we find that the average position $\alpha^*$ of the
boundary is given by the rate equation
\begin{equation} 2\alpha^* r_e = (1 - \alpha^*) r_e,
\label{rate}
\end{equation}
i.e. $\alpha^* = 1/3$. Interestingly, for not too high error rates $r_e$,  $\alpha^*$
is independent from $r_e$ and thus from $p$. If we consider the average boundary position  $\alpha^*$ as
a system-specific \emph{order parameter} which is controlled by the two quasi-particles,
then comparing the solution of Eqn.\ (\ref{rate}) to the equilibrium position in the noiseless case indicates
that the system
undergoes a first order phase transition with respect to  $\alpha^*$ at $p = r_e = 0$.
\begin{figure}[htb]
\let\picnaturalsize=N
\def\picsize{85mm}
\def\picfilename{fig8.eps}
\ifx\nopictures Y\else{\ifx\epsfloaded Y\else\input epsf \fi
\let\epsfloaded=Y
\centerline{\ifx\picnaturalsize N\epsfxsize \picsize\fi
\epsfbox{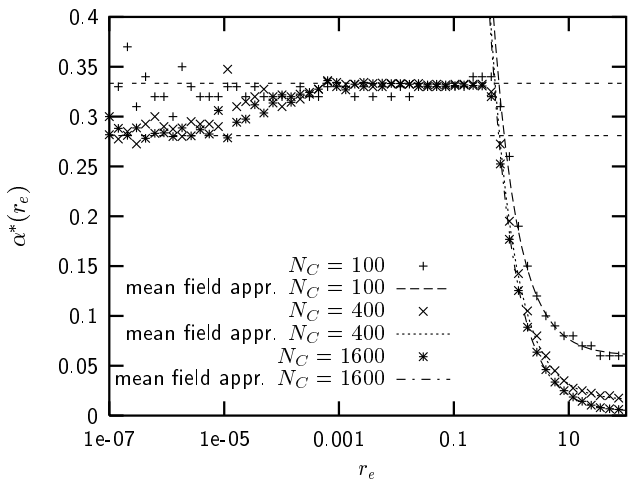}}}\fi
\caption{\small Average boundary position $\alpha^*$ as a function of the error rate $r_e$ for system sizes
$N_C = 100$, $N_C = 400$, and $N_C = 1600$. The abscissa is logarithmic. Numerical data are averaged over $200$ different initial conditions with
$2\cdot10^6$ updates each. The dashed curves show the mean field approximation given by Eqn.\ (5), the straight dashed lines
mark the unperturbed solution $\alpha^* = 0.281$ and the solution under noise, $\alpha^* = 1/3$. }
\label{phase}
\end{figure}
\begin{figure}[htb]
\let\picnaturalsize=N
\def\picsize{85mm}
\def\picfilename{fig9.eps}
\ifx\nopictures Y\else{\ifx\epsfloaded Y\else\input epsf \fi
\let\epsfloaded=Y
\centerline{\ifx\picnaturalsize N\epsfxsize \picsize\fi
\epsfbox{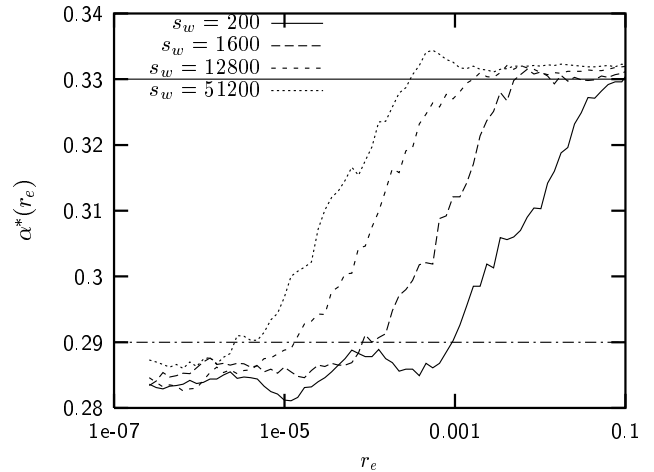}}}\fi
\caption{\small Average domain boundary position $\alpha^*$ as a function of the error rate $r_e$, sampled
over update windows of different lengths $s_w$ (ensemble statistics, 400 different initial conditions
for each data point). The abscissa is logarithmic.
With increasing $s_w$, the transition from the solution
$\alpha^*_{det} = 0.281$ under deterministic dynamics to $\alpha^* = 1/3$ under noise 
is shifted towards $r_e = 0$. The two straight lines define a lower boundary $\alpha^*_{low}$ and 
a upper boundary $\alpha^*_{up}$, as explained in the text.}
\label{pt0Shift}
\end{figure}
\begin{figure}[htb]
\let\picnaturalsize=N
\def\picsize{85mm}
\def\picfilename{fig10.eps}
\ifx\nopictures Y\else{\ifx\epsfloaded Y\else\input epsf \fi
\let\epsfloaded=Y
\centerline{\ifx\picnaturalsize N\epsfxsize \picsize\fi
\epsfbox{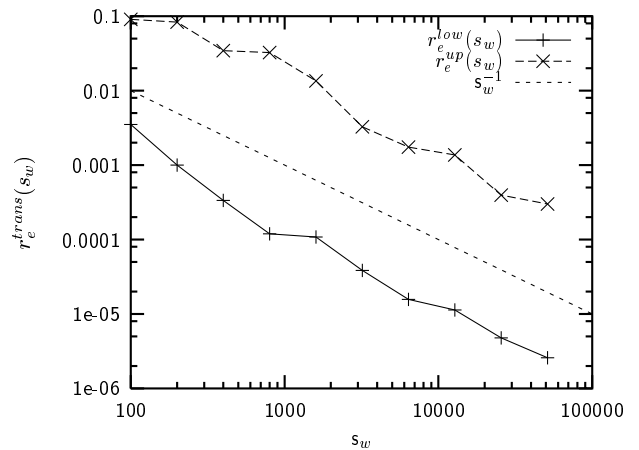}}}\fi
\caption{\small Finite size scaling of the upper and lower transition points $r_e^{up}$ and $r_e^{low}$,
i.e. the points where $\alpha^*$ crosses $\alpha^*_{low}$ and $\alpha^*_{up}$, respectively (Fig. 7),
as a function of the sampling window length $s_w$.
Both  $r_e^{up}$ and $r_e^{low}$ vanish $\propto s_w^{-1}$, as indicated by the line
with slope $-1$ in this log-log-plot. }
\label{pt0Scal}
\end{figure}

Figs.\ \ref{pt0Shift} and \ref{pt0Scal} show noise dependence and finite size scaling of the transition. 
In case of a first order phase transition at $r_e = 0$,
we would expect a shift of the transition point $r_e^{trans}(s_w)$ towards $r_e = 0$ which is proportional to $s_w^{-1}$ as well as 
a divergence of the slope at the transition point when $s_w$ is increased,  i.e. $d\alpha^*/dr_e(r_e^{trans}) \to \infty$ when $s_w \to \infty$.

The shift of $r_e^{trans}(s_w)$ is most easily measured by defining a lower and a upper boundary
$\alpha^*_{low}$ and $\alpha^*_{up}$, respectively (Fig. \ref{pt0Shift}); when  $\alpha^*$ crosses these boundaries,
two transition points $r_e^{up}$ and $r_e^{low}$ are obtained.
We find that $r_e^{up} \approx c_{up}s_w^{-1}$ and  $r_e^{low} \approx c_{low}s_w^{-1}$ with $c_{up} > c_{low}$ as
expected (Fig. \ref{pt0Scal}), which implies that the difference $\Delta r_e^{trans}(s_w) := r_e^{up} - r_e^{low}$ scales as
\begin{equation} \Delta r_e^{trans}(s_w) = (c_{up} - c_{low})\,s_w^{-1}, \end{equation} 
hence, because $\Delta \alpha^*(r_e^{trans}) = const. = \alpha^*_{up}- \alpha^*_{low}$,
indeed  $d\alpha^*/dr_e(r_e^{trans})$ diverges when the sampling window size goes to infinity. 

The solution  $\alpha^* = 1/3$ is stable only for $0 < r_e \le 1/2$. 
As shown in Fig.\ \ref{particles}, the interaction of a  $\Gamma$ particle with the boundary
needs only one update time step, whereas the boundary readjustment following a $\Delta$ particle 
interaction takes three update time steps. Hence, we conclude that the term on the right hand side of Eqn.\ (10), 
which gives the flow rate of $\Delta$ particles at the boundary, for large $r_e$ will saturate at $1/3$, leading to
\begin{equation} 
2\alpha^* r_e = \frac{1}{3} 
\end{equation} 
with the solution
\begin{equation} \alpha^*  = \frac{1}{6}\,\, r_e^{-1} + \Theta(N_C) \end{equation}
for $r_e > 1/2$. Hence, there is a crossover from the solution $\alpha^* = 1/3$ to another solution vanishing
with $r_e^{-1}$ around  $r_e = 1/2$.
The finite size scaling term $\Theta(N_C)$ can be estimated from the following
consideration: for $p \rightarrow 1$, the average domain size created by ``pure chance''
is given by $\alpha^* = N_C^{-1} \,\sum_{n=0}^{N_C} (1/3)^n \cdot n \approx (3/4)\,N_C^{-1}$. If the measuring
window has size $w$, we obtain $\Theta(N_C) \approx (3/4)\,w\,N_C^{-1}$.
To summarize, we find that the self-organized boundary position is given by
\begin{equation}
\alpha^*=  \left\{  
\begin{array}{cccc} 0.281 \pm 0.001  \quad &\mbox{if}& r_e =0 \\
 1/3  \quad &\mbox{if}& 0 < r_e \le 1/2 \\
         (1/6)r_e^{-1} + \Theta(N_C)  \quad &\mbox{if}&  \quad r_e > 1/2 &
\end{array} \right. 
\end{equation}
with a first-order phase transition at $r_e = 0$ and a crossover around $r_e = 1/2$. 

\begin{figure}[tb]
\resizebox{85mm}{!}{\includegraphics{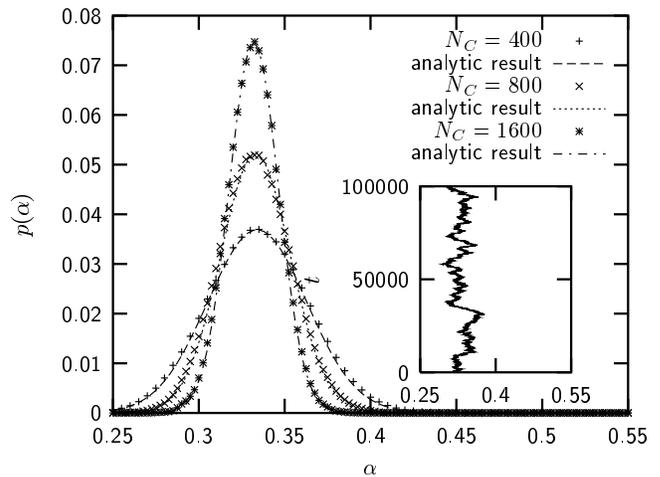}}
\caption{\small For the system with stochastic update errors, fluctuations of the boundary position $\alpha$
around the average position $\alpha^* = 1/3$ are Gaussian distributed. The figure compares the numerically
obtained stationary probability distribution with the analytic result of Eqn.\ (18) for three different system sizes. All data are gained for
$r_e = 0.1$ and averaged over $100$ different initial conditions with $2\cdot 10^6$ updates each. The inset shows a typical timeseries
of the boundary position. }
\label{fluct}
\end{figure}

Now let us consider the fluctuations of $\alpha$ around $\alpha^*$ given by the master equation
\begin{eqnarray} p^{\tau}(\alpha) &=& 2\alpha\,r_e\, p^{\tau-1}(\alpha + 2\delta) + (1-\alpha)\,r_e\, p^{\tau-1}(\alpha - \delta) \nonumber  \\
&+& (N_C-r_e)\, p^{\tau-1}(\alpha) - 2\alpha\,r_e\, p^{\tau-1}(\alpha) \nonumber \\  &-& (1-\alpha)\,r_e\, p^{\tau-1}(\alpha)
\end{eqnarray} 
with $\delta = 1/N_C$. Eqn.\ (15) determines the probability $p^{\tau}(\alpha)$ to find the boundary at position $\alpha$
at update time step $\tau$, given its position at time $\tau-1$. This equation can be simplified as we are interested only
in the \emph{stationary probability distribution} of $\alpha$. It is easy to see that the  error rate $r_e$ just provides 
a time scale for relaxation towards
the stationary distribution and has no effect on the stationary distribution itself. Therefore, we may consider the limit $r_e \to r_e^{max} := N_C$,
divide through $r_e$ and neglect the last three terms on the right handside of Eqn.\ (14) (which become
zero in this limit). We obtain 
\begin{equation} p^{\tau}(\alpha) = 2\alpha\, p^{\tau-1}(\alpha + 2\delta)  + (1-\alpha)\,  p^{\tau-1}(\alpha - \delta).  \end{equation} 
To study this equation, we consider the continuum limit $N_C \to \infty$.
Let us introduce the scaling variables $x = (\alpha - \alpha^*)\sqrt{N_C}$, $t = \tau /N_C$ and the probability density
$f(x,t) = N_C\, p^{\tau}(\alpha\, N_C)$. Inserting these definitions into Eqn.\ (16) and ignoring all subdominant powers $\mathcal{O}(1/N_C)$,
we obtain a Fokker-Planck equation:
\begin{equation} \frac{\partial f(x,t)}{\partial t} = \left( \frac{\partial^2}{\partial x^2} + 3 \frac{\partial}{\partial x}\,x\right) f(x,t).
 \end{equation} 
The stationary solution of this equation is given by
\begin{equation} f(x) = \sqrt{\frac{3}{2\pi}}\exp{\left[-\frac{3}{2}x^2 \right]},
 \end{equation} 
i.e. in the long time limit $t \to \infty$, the probability density for the boundary position $\alpha$ is a Gaussian with mean $\alpha^*$:
\begin{equation} p(\alpha, N_C ) = \sqrt{\frac{3\,N_C}{2\pi}}\exp{\left[-\frac{3\,N_C}{2}\left(\alpha-\alpha^*\right)^2 \right]}.
 \end{equation}
From Eqn.\ (18) we see that the variance of $\alpha$ vanishes $\sim 1/N_C$ and the relative boundary position becomes sharp
in the limit of large system sizes. Fig. \ref{fluct} shows that this continuum approximation for $N_C \ge 400$ provides
very good correspondence with the numerically obtained probability distributions.

\begin{figure}[htb]
\let\picnaturalsize=N
\def\picsize{85mm}
\def\picfilename{fig12.eps}
\ifx\nopictures Y\else{\ifx\epsfloaded Y\else\input epsf \fi
\let\epsfloaded=Y
\centerline{\ifx\picnaturalsize N\epsfxsize \picsize\fi
\epsfbox{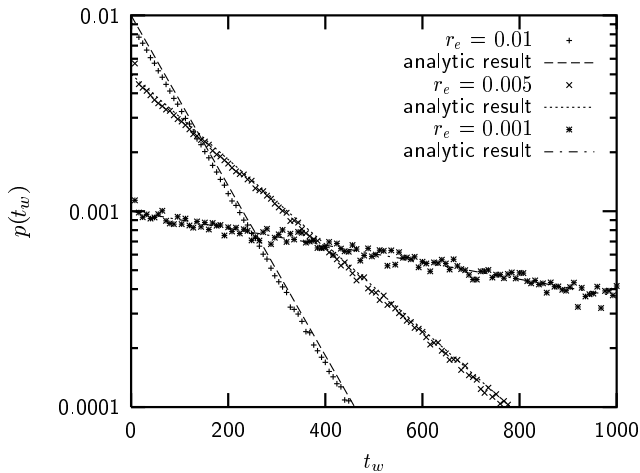}}}\fi
\caption{\small Probability distribution $p(t)$ of waiting times for boundary readjustements in the model
with stochastic update errors for three different values of $r_e$, semi-log plot. As expected for a Poisson process,
$p(t)$ is an exponential. }
\label{wait}
\end{figure}

The stochastic nature of boundary stabilization under noise is also reflected
by the probability distribution of \emph{waiting times} $t$ for boundary readjustments
due to particle interactions: the particle production is a Poisson process with
the parameter $\lambda = r_e$ and the waiting time distribution is given by
\begin{equation} p_{wait}(t) =  r_e\exp{(-r_e\,t)} \end{equation}
with an average waiting time $\langle t \rangle = r_e^{-2}$. Fig.\ \ref{wait} shows the waiting time
distributions for different error rates $r_e$.
In a biological organism, a pattern has to be robust not only with respect
to dynamical noise, but also with respect, e.g., to ``mechanical'' perturbations. In \emph{Hydra}, e.g., 
there is a steady flow of cells directed towards the animal's head and foot, due to continued proliferation
of stem cells \cite{DavidCamp72}; the stationary pattern of gene activity
is maintained is spite of this cell flow. Let us now study the robustness of the model with respect to this type of perturbation.
Let us consider a constant cell flow with rate $r_f$, which is directed
towards the left or the right system boundary. In Eqn.\ (9), we now get an additional drift term $r_f$
on the left hand side:
\begin{equation} 2\alpha^*r_e \pm  r_f = r_e(1 - \alpha^*),  \label{20}\end{equation} 
with the solution
\begin{equation}
\alpha^*=  \left\{  \begin{array}{cccc} \frac{1}{3}\left ( 1 - \frac{r_f}{r_e}\right )  \quad &\mbox{if}&  r_e \ge r_f \\ 
                                          0     \quad  &\mbox{if}&  r_e < r_f &
\end{array} \right. 
\end{equation}
for the case of cell flow directed towards the left system boundary (plus sign in eqn. (\ref{20})).
One observes that $\alpha^*$ undergoes a second order phase transition at the critical value $r_e^{crit} = r_f$. 
Below $r_e^{crit}$, the domain size $\alpha^*$ vanishes, and above $r_e^{crit}$ it grows until it reaches
the value $\alpha^*_{max}=1/3$ of the system without cell flow. For cell flow directed
towards the right system boundary (minus sign in eqn. (\ref{20})), we obtain
\begin{equation}
\alpha^*=  \left\{  \begin{array}{cccc} 
\frac{1}{3}\left ( 1 + \frac{r_f}{r_e}\right )  \quad &\mbox{if}&  r_f \le 2\,r_e \\ 
                   1     \quad  &\mbox{if}&  r_f > 2\,r_e.
\end{array} \right. 
\end{equation}
In this case, the critical cell flow rate is given by $r_f = 2\, r_e$, for cell flow rates larger
than this value the $\sigma_2$-domain extends over the whole system, i.e.\ $\alpha^* = 1$. 
\begin{figure}[htb]
\let\picnaturalsize=N
\def\picsize{85mm}
\def\picfilename{fig13.eps}
\ifx\nopictures Y\else{\ifx\epsfloaded Y\else\input epsf \fi
\let\epsfloaded=Y
\centerline{\ifx\picnaturalsize N\epsfxsize \picsize\fi
\epsfbox{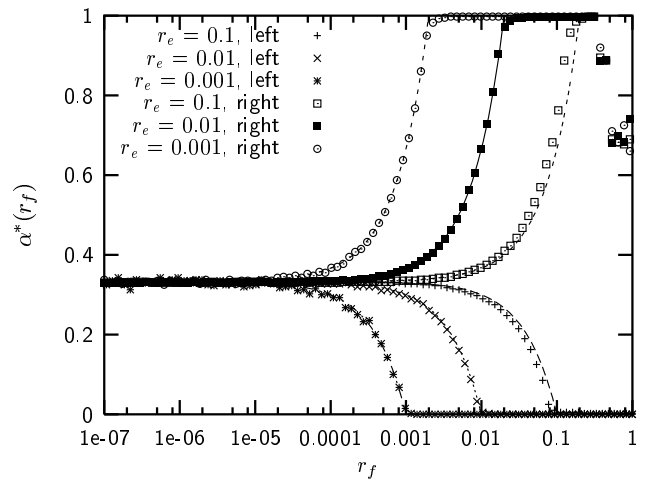}}}\fi
\caption{\small Average domain size $\alpha^*$ as a function of the the cell flow rate $r_f$ for three
different error rates $r_e$; numerical data (crosses and points) were sampled over 10 different initial conditions and 1e6
updates for each data point. ``Left'' indicates cell flow directed to the left system boundary, ``right'' to the right system boundary,
respectively. The dashed lines are the corresponding solutions of Eqn.\ (\ref{20}).}
\label{cnflow}
\end{figure}
Fig.\ \ref{cnflow} compares the results of numerical simulations
with the mean field approximation of Eqn.\ (\ref{20}). In numerical simulations, cell flow is realized by application
of the translation operator $\Theta\,\sigma_i := \sigma_{i+1}$ to all cells with $0 \le i < N_C-1$ every $r_f^{-1}$ time steps
and leaving $\sigma_{N_C-1}$ unchanged. In case of cell flow directed to the right system boundary, in the limit
$r_f \to 1$ the boundary position $\alpha^*$ detected in numerical simulations deviates from the mean field prediction,
due to a boundary effect at the left system boundary (stochastic production of finite lifetime stationary oscillators, leading
to intermittent flows of $\Gamma$ particles through the system).

To summarize this part, we see that in the model stochastic errors in dynamical updates for $r_e > r_f$ indeed \emph{stabilize} the global pattern
against the mechanical stress of directed cell flow.

\section{Summary and Discussion}
In this paper we considered the dynamics of pattern formation motivated by animal morphogenesis 
and the largely observed participation of complex gene regulation networks in their coordination and control. 
We therefore chose a simple developmental problem to study toy models of interacting networks that 
control pattern formation and morphogenesis in this setting. In particular, the goal was to explore 
how networks can offer additional mechanisms beyond the standard diffusion based process of the 
Turing instability. Our results suggest that main functions of morphogenesis can be performed by 
dynamical networks without relying on diffusive biochemical signals, but using local signaling between 
neighboring cells. This includes solving the problem of generating global position information from purely
local interactions. But also it goes beyond diffusion based models as it offers solutions to developmental 
problems that are difficult for such models and avoids their inherent problem of fine tuned model parameters. 
It may be not too far fetched to speculate that some developmental processes as, e.g., the establishment
of position information, may rely on this type of internal information processing rather than on interpretation 
of, e.g., chemical gradients. As only ingredient to the present model, symmetry has to be
broken locally by a spatially asymmetric receptor distribution, e.g., as proposed in \cite{Marciniac03}, 
to provide the necessary input information. Also hybrid approaches are conceivable. For example such 
broken symmetry could be provided by a gradient, even without fine-tuning (contrary to standard gradient-based models) 
since only the direction of the gradient has to be read out.  

The network model derived here performs accurate regulation of position information and robust {\em de novo} pattern formation
from random conditions, with a mechanism based on local information transfer rather than the Turing instability. 
Non-local information is transmitted through soliton-like quasi-particles instead of long-range gradients.
Two realizations as discrete dynamical networks, Boolean networks and threshold networks, have been 
developed. The resulting networks have size and complexity comparable to developmental gene regulation modules as
observed in animals, e.g., \emph{Drosophila} \cite{vonDassow} or \emph{Hydra} \cite{BoschDevelGenes}. 
The threshold networks (as models for transcriptional regulation networks) process position information in
a hierarchical manner; in the present study, hierarchy levels were limited to three, but realizations with more
levels of hierarchy, i.e.\ more ``pre-processing'' of information are also possible. Similar hierarchical and modular organization are
typical signatures of gene regulatory networks in organisms \cite{Davidson01}. 

Robustness of the model is studied in detail for two types of perturbations, stochastic update errors (noise)
and directed cell flow.  A first order phase transition is observed for vanishing noise and a second order phase transition 
at increasing cell flow. Fluctuations of the noise-controlled boundary position were studied numerically for finite size
systems and analytically in the continuum limit. We find that the relative size of fluctuations vanishes with $1/N_C$,
which means that the boundary position becomes sharp in the limit of large system sizes. Dynamics under cell flow
is studied in detail numerically and analytically by a mean field approximation. A basic observation is that 
noise-induced perturbations act as quasi-particles that 
stabilize the pattern against the directed force of cell flow. At a critical cell flow rate, there is a second order
phase transition towards a vanishing domain size or a domain extending over the whole system, depending on the direction
of cell flow, respectively. 

Several extensions of this model are conceivable. 
In the present model, the cell flow rate $r_f$ is considered as a free parameter, the global pattern, however, can be controlled
easily by an appropriate choice of the error rate $r_e$. This may suggest to extend the model by introduction of some kind of
dynamical coupling between $r_e$ and $r_f$, treating $r_f$ as a function of $r_e$. Interestingly, similar approaches have been
studied by Hogeweg \cite{Hogeweg2000} and Furusawa and Kaneko \cite{Furusawa2000,Furusawa2003}: 
In both models of morphogenesis, the rate of cell divisions is
controlled by cell differentiation and cell-to-cell signaling. Dynamics in both models, however, is deterministic.
An extension of our model as outlined above may open up for interesting studies how {\em stochastic} signaling events could
control and stabilize a global expression pattern and cell flow as an integrated system. Other possible extensions of the model concern
the dimensionality: In two or three dimensions other mechanisms of symmetry breaking might be present, possibly
leading to new, interesting dynamical effects. 

A Java applet simulation of the model can be found at \cite{Applet}.

\section{Acknowledgements}
We thank T.C.G.\ Bosch, T.W.\ Holstein, and U.\ Technau  
for pointing us to current questions in Hydra development. 
T.R.\ acknowledges financial support from 
the Studienstiftung des deutschen Volkes (German National Merit Foundation).


\begin{appendix}

\section{Genetic algorithm searches}
Let us briefly recapitulate here how the rule table of the model has been obtained by the aid of a 
genetic algorithm. 
\subsection{Definition of the GA}
In order to find a set $\cal F$ of update rules that solve the problem as formulated in section II, 
cellular automata have been evolved using genetic algorithms \cite{Mitchell94}. Genetic algorithms are 
population-based search algorithms, which are inspired by the interplay of random mutations 
and selection as observed in biological evolution \cite{Holland}. Starting from a randomly 
generated population of $P$ rule tables $f_n$, the algorithm optimizes possible solutions by
evaluating the fitness function
\begin{eqnarray}
\Phi(f_n) = \frac{1}{(T_u-T)\cdot N_C}\sum_{t = T_u-T}^{T_u}\left(\sum_{i=0}^{[\alpha\cdot N_C]-1}\delta_{\sigma_i^n(t),2} \right. \nonumber\\
        \left. + \sum_{i=[\alpha\cdot N_C]}^{N_C-1}\{1-\delta_{\sigma_i^n(t),2}\} \right).
\end{eqnarray}
The optimization algorithm then is defined as follows: 
\begin{enumerate}
\item Generate a random initial population $\cal F$ $= \{ f_1,...,f_P \}$ of rule tables.
\item Randomly assign system sizes $N_C^{min} \le N_c^n \le N_C^{max}$ to all rule tables.
\item For each rule table, generate a random initial state vector.
\item Randomly mutate one entry of each rule table (generating a population $\cal F^*$ of mutants).
\item Iterate dynamics over $T_u$ time steps for $\cal F$ and  $\cal F^*$. 
\item Evaluate $\Phi(f_n)$ and $\Phi(f_n^*)$ for all rule tables $0 < n \le P$, averaging over the past 
$T_u-T$ update steps (with an additional penalty term
      if $f_n$ does not converge to a fixed point).
\item For each $n$, replace $f_n$ with $f_n^*$, if $\Phi(f_n) \le \Phi(f_n^*)$.
\item Replace the least fit solution by a duplicate of the fittest one.
\item Go back to step 2 and iterate.
\end{enumerate}
The outcome of this search algorithm is a set of rule tables, which then can be ``translated''
into (spatially coupled) Boolean networks or threshold networks with suitable thresholds.
This yields a set of (minimal) dynamical networks which solve the pattern formation
task by means of internal information processing.
\begin{figure}[htb]
\let\picnaturalsize=N
\def\picsize{85mm}
\def\picfilename{fig14.eps}
\ifx\nopictures Y\else{\ifx\epsfloaded Y\else\input epsf \fi
\let\epsfloaded=Y
\centerline{\ifx\picnaturalsize N\epsfxsize \picsize\fi
\epsfbox{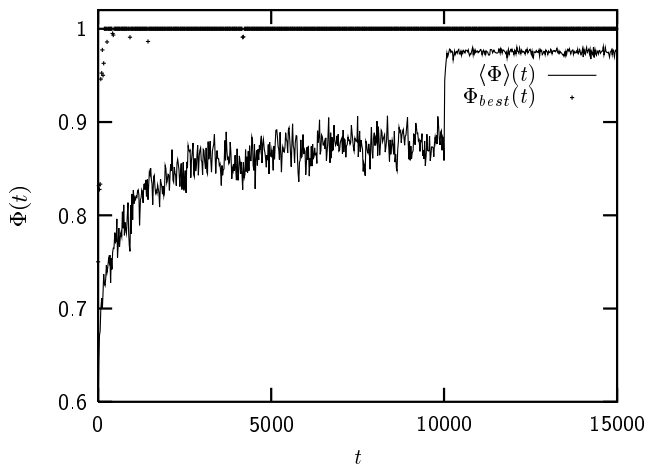}}}\fi
\caption{\small Average fitness $\langle \Phi\rangle(t)$ of the mutant population $\cal F^*$ 
and fitness of the highest fitness mutant $\Phi_{best}(t)$ as a function of simulation time 
during the genetic algorithm run that lead to the high fitness solution used in this paper.
At time step 10000 mutations
were turned off, in order to test the established population of optimized rule tables under different
initial conditions (this corresponds to the sharp increase of  $\langle \Phi\rangle(t)$ at time step 10000). 
The evolved population of rule tables has an average fitness of about 0.98, independent from the initial conditions and
system size $N_C$ (in the tested range, i.e. $15 \le N_C \le 150$).}
\label{evo}
\end{figure}
\begin{figure}[htb]
\let\picnaturalsize=N
\def\picsize{85mm}
\def\picfilename{fig15.eps}
\ifx\nopictures Y\else{\ifx\epsfloaded Y\else\input epsf \fi
\let\epsfloaded=Y
\centerline{\ifx\picnaturalsize N\epsfxsize \picsize\fi
\epsfbox{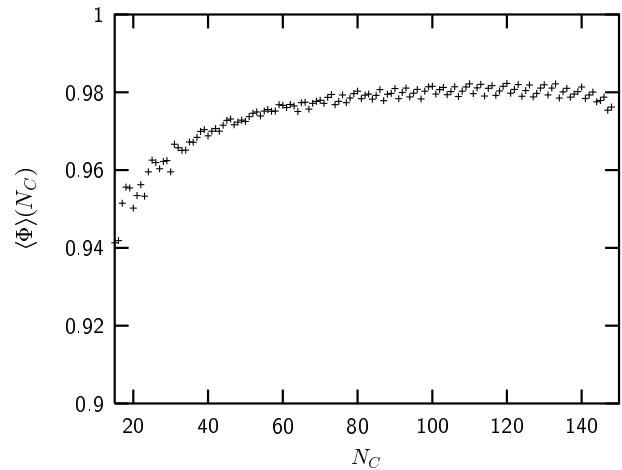}}}\fi
\caption{\small Average fitness of the highest fitness rule table as a function of the system size $N_C$. For system sizes
$N_C \ge 80$ the fitness is almost constant at about 0.98. Notice that the decrease of the fitness for small
$N_C$ is an effect of the \emph{dynamics}, not of the genetic algorithm implementation (all $N_C$ in the range $15 \le N_C\le 150$
were tested with equal probability), hence the dynamics of pattern formation may impose a lower boundary
on the range of system (animal) sizes tolerated by natural selection.  }
\label{fitOfN}
\end{figure}
\begin{figure}[htb]
\let\picnaturalsize=N
\def\picsize{80mm}
\def\picfilename{fig16.eps}
\ifx\nopictures Y\else{\ifx\epsfloaded Y\else\input epsf \fi
\let\epsfloaded=Y
\centerline{\ifx\picnaturalsize N\epsfxsize \picsize\fi
\epsfbox{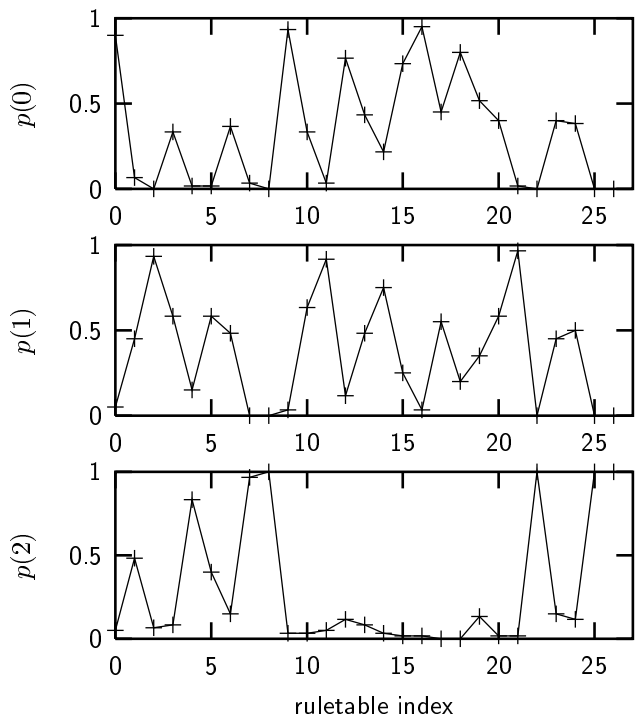}}}\fi
\caption{\small Frequency distribution $p(\sigma_i)$ of outputs as a function of the rule table
index as denoted in table 1. Ensemble statistics is taken
over 80 different solutions with $\Phi \ge 0.96$. The upper panel shows the distribution for
$\sigma_i = 0$, the middle panel  the distribution for $\sigma_i = 1$ and the lower panel
 the distribution for $\sigma_i = 2$.}
\label{rulstat}
\end{figure}
\begin{figure}[tb]
\resizebox{85mm}{!}{\includegraphics{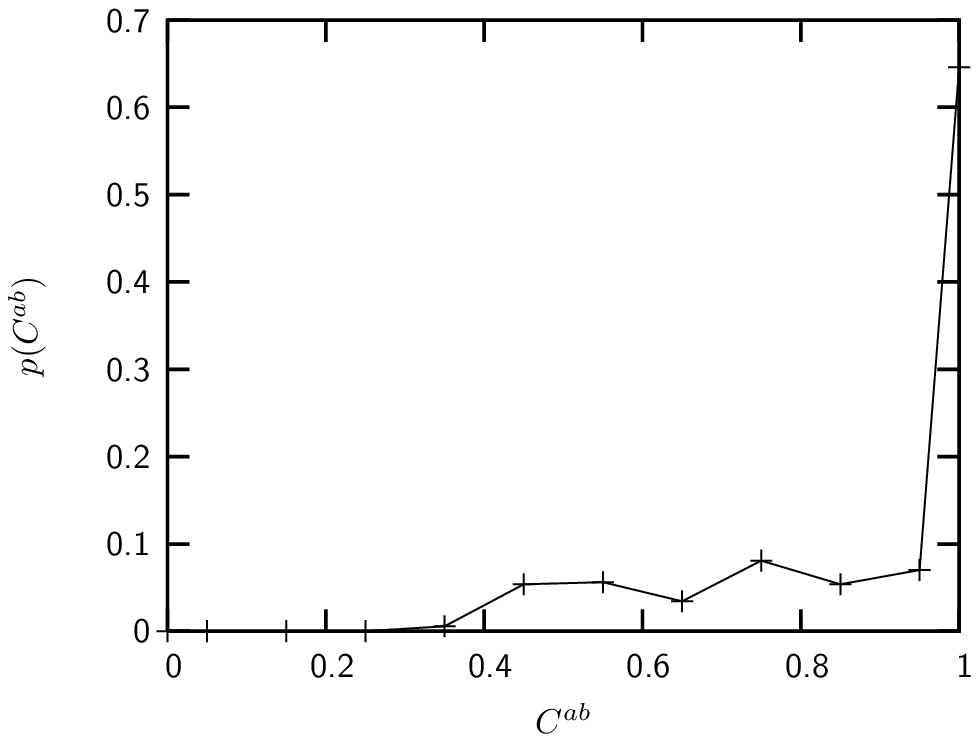}}
\caption{\small Frequency distribution $p(C^{ab})$ of two point correlations $C^{ab}(\sigma,\sigma^{'})$ 
of rule table entries, as defined in Eqn.\ (B2), averaged over all possible pairs of rule table entries. 
About 65\% of rule table entries have correlation 1.0, for the rest the correlation is lower.}
\label{rulcorr}
\end{figure}

\subsection{Evolution of cellular automata}

The genetic algorithm sketched above is run with the following parameter
choices: $15 \le N_C \le 150$, i.e.\ during GA runs the system size is 
varied randomly between $15$ and $150$ cells, and the population size is set to
$P = 100$.
 Fig.\ \ref{evo} shows the fitness of the highest fitness mutant and the average
fitness of the population as a function of the number of successive mutation
steps during optimization. A useful solution is found rather quickly (after about
200 updates), with further optimization observed during futher 10000 generations. 
At time step 10000 mutations are turned off, thus now
the average fitness of the established population under random initial conditions
and random fluctuations of the system size $N_C$ is tested. The average
fitness $\Phi \approx 0.98$ indicates a surprisingly high robustness
against fluctuations in the initial start pattern, indicating that the system
is capable of \emph{de novo pattern formation}.
In the ``fitness picture'', Fig. \ref{fitOfN} confirms
that the dynamically regulated domain size ratio $\alpha/(1-\alpha)$
indeed is independent of system size (proportion regulation),
there is only a weak decay of the fitness at small values of $N_C$.
Interestingly, one also observes that for regeneration of Hydra polyps
from random cell aggregates a minimum number of cells is required
\cite{TechnauHolstein}. The model suggests that this observation might be explained
by the dynamics of an underlying pattern generating mechanism, i.e.\ that there
has to be a minimum diversity in the initial condition for successful
de novo pattern formation.

\section{Statistical analysis of solutions}

An interesting question is how ``difficult'' it would be for an evolutionary 
process driven by random mutations and selection to find solutions
for the pattern formation problem based on neighbor interactions
between cells. As we showed above, the genetic algorithm finds the correct solution
fast, however, this does not necessarily
mean that biological evolution could access the same solution as fast. If there
is only one, singular solution, evolution may never succeed finding it,
as the genotype which already exists cannot be modified in an arbitrary
way without possibly destroying function of the organism (\emph{developmental
constrains}). To illustrate this point, we generated an ensemble of $N_E = 80$ different solutions
with $\Phi \ge 0.96$ and performed a statistical analysis of the rule table
structure.

As one can see in Fig.\ \ref{rulstat}, some positions in the rule table
are quite fixed, i.e., there is not much variety in the outputs,
whereas other positions are more variable. However, the output frequency
distribution alone does not allow to really judge the ``evolvability''
of the solutions: if there are strong correlations between most of the
rule table entries, evolutionary transitions from one solution to another
would be almost impossible. To check this point,
we studied statistical two point correlations between the rule table entries. The 
probability for finding state $\sigma$
at position $a$ and state $\sigma'$ at rule table position $b$ is given by 
\begin{equation} p^{ab}(\sigma,\sigma') = \frac{1}{N_E}\sum_{n=1}^{N_E}\delta_{\sigma^n(a),\sigma}\cdot\delta_{\sigma^n(b),\sigma'}
,\end{equation}
where $\delta$ is the Kronecker symbol and $n$ runs over the statistical ensemble 
of size $N_E$. The two point correlation between $a$ and $b$ then is defined as 
\begin{equation} C^{ab} = c_1\left( \max_{(\sigma, \sigma')} p^{ab}(\sigma,\sigma') - c_2\right) \end{equation}
with $c_1 = 9/8$ and $c_2 = 1/9$ to obtain a proper normalization with respect to the two limiting cases of equal
probabilities ($p^{ab}(\sigma,\sigma{'}) = 1/9 \quad \forall (\sigma,\sigma{'})$) and $p^{ab}(\sigma,\sigma{'}) = 1$ for 
$\sigma = \tilde\sigma, \quad \sigma^{'} = \tilde\sigma^{'}$
and $p^{ab}(\sigma,\sigma') = 0$ for all other $(\sigma,\sigma')$).
Fig.\ \ref{rulcorr} shows the frequency distribution of $C^{ab}(\sigma,\sigma')$, 
averaged over all possible pairs $(a,b)$. About 65\% of rule table positions 
are strongly correlated ($C^{ab} = 1.0$), the rest shows correlation values
between $0.3$ and $1.0$. Hence, we find that the space of solutions is 
restricted, nevertheless there is variability in several rule table positions. 
To summarize this aspect, the pattern formation mechanism studied in this 
paper shows considerable robustness against rule mutations, however,  
a ``core module'' of rules is always fixed. Interestingly, a similar phenomenon 
is observed in developmental biology: Regulatory ``modules'' involved 
in developmental processes often are evolutionaryly very conservative,
 i.e., they are shared by almost all animal phyla \cite{Davidson01}, while 
 morphological variety is created by (few) taxon specific genes 
 \cite{BoschDevelGenes} and ``rewiring'' of existing developmental modules.

\end{appendix}

\end{document}